\documentstyle[12pt,epsf]{article}
\newcommand{\postscript}[2]{\setlength{\epsfxsize}{#2\hsize}
   \centerline{\epsfbox{#1}}}
\def\be{\begin{equation}}
\def\ee{\end{equation}}
\def\bear{\begin{eqnarray}}
\def\eear{\end{eqnarray}}

\def\NPB#1#2#3{{\it Nucl.~Phys.} {\bf{B#1}} (19#2) #3}
\def\PLB#1#2#3{{\it Phys.~Lett.} {\bf{B#1}} (19#2) #3}
\def\PRD#1#2#3{{\it Phys.~Rev.} {\bf{D#1}} (19#2) #3}
\def\PRL#1#2#3{{\it Phys.~Rev.~Lett.} {\bf{#1}} (19#2) #3}
\def\ZPC#1#2#3{{\it Z.~Phys.} {\bf C#1} (19#2) #3}
\def\PTP#1#2#3{{\it Prog.~Theor.~Phys.} {\bf#1}  (19#2) #3}

\def\MPLA#1#2#3{{\it Mod.~Phys.~Lett.} {\bf A#1} (19#2) #3}

\def\EPJ#1#2#3{{\it Eur.~Phys.~J.} {\bf C#1} (19#2) #3}

\def\roughly#1{\raise.3ex\hbox{$#1$\kern-.75em\lower1ex\hbox{$\sim$}}}
\newcommand{\dr}{\mbox{\footnotesize$\overline{\rm DR}$~}}
\newcommand{\ms}{\mbox{\footnotesize$\overline{\rm MS}$~}}
\newcommand{\olf}{16\pi^2}
\newcommand{\tAt}{X_t}
\newcommand{\htAt}{{\widehat{X}}_t}

\newcommand{\tilt}{\tilde{t}}

\newcommand{\mtilt}{m_{\tilde{t}}}
\newcommand{\MQ}{M_{\widetilde{Q}}}
\newcommand{\MU}{M_{\widetilde{U}}}
\newcommand{\hmst}{{\widehat{m}}_{\tilde{t}}}
\newcommand{\bmt}{\overline{m}_t}
\newcommand{\lntt}{\ln{\mtilt^2\over m^2_t}}
\newcommand{\lntq}{\ln{m_t^2\over Q^2}}
\newcommand{\lnttq}{\ln{\mtilt^2\over Q^2}}
\newcommand{\gsim}{\lower.7ex\hbox{$\;\stackrel{\textstyle>}{\sim}\;$}}
\newcommand{\lsim}{\lower.7ex\hbox{$\;\stackrel{\textstyle<}{\sim}\;$}}

\relax

\begin{document}
\begin{titlepage}
\begin{flushright}
IFT-UAM/CSIC-99-47\\
MADPH-99-1149\\
hep-ph/9912236 \\
\end{flushright}
\vskip 0.3in
\begin{center}{\Large\bf
MSSM Lightest CP-Even 
Higgs Boson Mass to ${\cal O}(\alpha_s\alpha_t)$:\\[2mm]
the Effective Potential Approach} 
\vskip .5in
{\bf Jose Ram\'on Espinosa}\\
IMAFF (CSIC)\\
Serrano 113 bis, 28006 Madrid, SPAIN\\
and \\
{\bf Ren-Jie Zhang}\\
Department of Physics, University of Wisconsin\\
1150 University Avenue, Madison Wisconsin 53706, USA\\

\end{center}
\vskip.5cm
\begin{center}
{\bf Abstract}
\end{center}
\begin{quote}
Starting with the two-loop effective potential of the MSSM, and assuming
a supersymmetric scale well above $M_Z$, we derive a simple analytical
approximation for the lightest CP-even Higgs boson mass including 
resummation of higher
order logarithmic terms via RG-improvement and
finite non-logarithmic terms up to ${\cal O}(\alpha_s\alpha_t)$. 
This formula describes the most
relevant radiative corrections to the MSSM Higgs boson mass, in particular,
those associated with non-zero top-squark mixing. 
\end{quote}
\vskip1.cm

\begin{flushleft}
December 1999 \\
\end{flushleft}

\end{titlepage}
\setcounter{footnote}{0}
\setcounter{page}{1}
\newpage
%
\noindent

\section{Introduction}

Supersymmetry (SUSY) stabilizes the hierarchy between the scale of electroweak
symmetry breaking and the fundamental energy scale (the GUT, Planck or
string scale) and makes plausible that this breaking occurs in a weakly
coupled Higgs sector. The simplest realistic model that accommodates these
ideas is the Minimal Supersymmetric Standard Model (MSSM), the
paradigmatic 
testing ground of low-energy SUSY.
Given a weakly coupled Higgs sector and a scale of electroweak
symmetry breaking $v=246$ GeV, it follows \cite{lh} that the
spectrum of the theory must contain an scalar particle $h^0$ with mass
controlled by the Fermi scale and non-zero couplings to the $W^\pm$ and
$Z^0$ gauge bosons (which is crucial for its detection at accelerators
\cite{eg}). In the MSSM, the mass of this light Higgs boson (besides this
Higgs particle, the Higgs sector in the MSSM contains a heavier scalar
$H^0$, one pseudoscalar $A^0$ and a pair of charged Higgses $H^\pm$) is
calculable. A precise determination of this mass is of paramount importance
for the experimental search of SUSY. If the experimental lower 
bound on the Higgs boson mass \cite{LEP2} increases above the theoretical
prediction
we can rule out the MSSM. It is therefore understandable that this
calculation has received a great deal of attention and attracted the
efforts of many groups \cite{rad}-\cite{HHWan}.

At tree level, the mass squared, $m_{h^0}^2$, of the light Higgs boson 
has an upper
bound (saturated for large values of the pseudoscalar mass $m_{A^0}$) given
by $M_Z^2\cos^22\beta$. This is already below the experimental lower bound
from LEP2 \cite{LEP2}. However, radiative corrections can raise the upper
bound on $m_{h^0}^2$ dramatically \cite{rad}. The dominant contribution is
\begin{equation}
\label{leading}
\Delta m_{h^0}^2\ =\ {3 m_t^4\over2\pi^2v^2}\ln{m_{\tilde{t}}^2\over m_t^2}\ ,
\end{equation}
where $m_t$ is the top quark mass and $m_{\tilde{t}}$ a common top-squark mass.
Radiative corrections to $m_{h^0}^2$ have been computed using different
techniques: effective potential approach, direct diagrammatic calculation 
and effective theories with renormalization group (RG) tools. Each approach
has its own virtues. The effective potential way simplifies the
computation; the diagrammatic calculation is unavoidable to pick up some
particular corrections; and the RG approach permits resummation of
logarithmic terms to all loops and provides a physically meaningful 
organizing principle for the radiative corrections. There is no real need
to choose one among these methods: the best way to proceed is to combine
the three methods, taking advantage of the best virtue of each in turn.

The $n$-loop contribution to the (dimensionless) quantity
$m_{h^0}^2/m_t^2$
has
the schematic form 
\be
\label{radstr}
\sum_{k=0}^n \left(\alpha \ln{m_{\tilde{t}}\over m_t}\right)^k \alpha^{n-k} ,
\ee
where $\alpha$ represents the expansion parameter  
(with $\alpha_t=h_t^2/4\pi$ and $\alpha_s=g_s^2/4\pi$ giving the
dominant contributions), and the logarithms $\ln(m_{\tilde{t}}/m_t)$
can be sizable for large values of $m_{\tilde{t}}$. The term $k=n$
in Eq.~(\ref{radstr}) is the leading-logarithmic correction and dominates
the 
$n$-loop contribution for large $m_{\tilde{t}}$. The term $k=n-1$ is the
sub-leading logarithmic term, etc. Finally, the term $k=0$ gives the finite
non-logarithmic piece of the $n$-loop correction.
The complete one-loop radiative corrections to $m_{h^0}^2/m_t^2$ have been 
calculated. The dominant leading-logarithmic 
contribution is the $\alpha_t \ln(m_{\tilde{t}}/m_t)$
term in Eq. (\ref{leading}). At this loop order there are no $\alpha_s$
corrections, which enter at two-loops. The most important finite
corrections depend on the top-squark mixing parameter
$\tAt=A_t+\mu\cot\beta$, is given by 
\be
\Delta m_{h^0}^2\ =\ {3m_t^4\over2\pi^2 v^2} 
\left({\tAt^2\over\mtilt^2}
-{\tAt^4\over12\mtilt^4}\right)\ ,
\ee
and its effect on $m_{h^0}^2$ can be
important. The maximum value for the upper bound on the Higgs mass is
obtained for $\tAt^2=6 m_{\tilde{t}}^2$ (the so-called
`maximal-mixing' case). 

The most important part of the higher order radiative corrections can be 
collected and resummed using renormalization group techniques
\cite{CEQWHHH}. Resumming with one-loop RG equations 
takes into account the
leading-logarithmic corrections to all loops, 
and using two-loop RG equations sub-leading
logarithms are also included. However, the finite non-logarithmic 
terms cannot be
obtained in this way. Getting them at two-loops requires a genuine two-loop
calculation. This was first done, in some particularly simple limit, in
Ref.~\cite{HH}, but the effect of non-zero top-squark mixing was not included.
Recently, a two-loop diagrammatic computation to  ${\cal
O}(\alpha_t\alpha_s)$ has
been completed \cite{HHW}, and the effective potential has been also
calculated to the same order \cite{RenJie}. These studies show that the
most significant two-loop effect, not previously taken into account by RG
techniques, comes from the finite pieces dependent on the top-squark mixing. The
new refined upper bound on $m_{h^0}$ in the region of maximal mixing can
increase up to 5 GeV (if $m_{\tilde{t}}\sim 1$ TeV) and the condition for
maximal-mixing itself gets also slightly modified.\footnote{This is not a
disagreement between different calculations, but the result of
including in the analysis of \cite{HHW} two-loop effects
inaccessible to previous approaches.}

Making good use of RG resummation, it is possible to obtain compact
analytical approximations for $m_{h^0}^2$ which take into account the most
important radiative corrections \cite{CEQWHHH}. Besides being of
direct practical interest, these formulae are theoretically interesting as
they provide a clear picture of the physical origin of the dominant
contributions. In this paper we extract such an analytical approximation
for $m_{h^0}^2$ starting with the two-loop effective potential computed in
\cite{RenJie}. Our final formula includes the most important ${\cal
O}(\alpha_s\alpha_t)$ radiative corrections, in particular the finite
terms associated with non-zero top-squark mixing. Our results agree with those
obtained by diagrammatic techniques \cite{HHWan} where they overlap, but 
are computed by an alternative way. We include, in addition, RG
resummation which allows us to write a particularly simple final formula
for the radiative corrections to $m_{h^0}^2$; 
this is the main result of our paper.
Again, to the order at which previous RG results \cite{CEQWHHH} were
computed, we find agreement with our results. Our final formula for
$m_{h^0}^2$ improves over previous RG formulae by including the genuine
two-loop threshold corrections (which are important for large values of
the top-squark mixing) and over previous diagrammatic results by
incorporating RG-resummation of logarithmic corrections.

\section{Two-loop effective potential in the MSSM}

The two-loop effective potential in the MSSM to  
${\cal O}(\alpha_s\alpha_t)$ has been calculated for the case of 
zero top-squark mixing and $\tan\beta=\infty$ in \cite{HH},
where $\tan\beta=v_2/v_1$ is the usual ratio of the vacuum expectation
values of the two Higgs fields;
and for arbitrary top-squark mixings and $\tan\beta$ 
values in \cite{RenJie}\footnote{
We choose to work in the $\dr$ scheme \cite{DR}, as required
by preserving the SUSY Ward identities. This choice also simplifies
complications associated with vector-boson loops in the $\ms$ scheme.
The two-loop effective potential in Eq. (\ref{2lep}) is 
calculated in the Laudau gauge, but it is 
also correct for a general $R_\xi$ gauge.},
\begin{eqnarray}
&&(\olf)^2~V_{2}(h_1,h_2)\ =\
8 g^2_3 \Biggl\{J(m_t,m_t)-2 m^2_t\ I(m_t,m_t,0) \nonumber\\
&& +{1\over 2}(c^4_t+s^4_t)\sum_{i=1}^2 J(m_{\tilde t_i},m_{\tilde t_i}) 
+ 2 s^2_t c^2_t J(m_{\tilde t_1},m_{\tilde t_2})
+ \sum_{i=1}^2 m^2_{\tilde t_i} I(m_{\tilde t_i},m_{\tilde t_i},0)\nonumber\\
&& +2J(m_{\tilde g},m_t) - \sum_{i=1}^2 \biggl[J(m_{\tilde t_i},m_{\tilde g})
+J(m_{\tilde t_i},m_t) + (m_{\tilde t_i}^2 - m^2_{\tilde g} - m^2_t)
I(m_{\tilde t_i},m_{\tilde g},m_t)\biggr]\nonumber\\
&&-4 m_{\tilde g}\ m_t\ s_t c_t
\left[I(m_{\tilde t_1},m_{\tilde g},m_t)
-I(m_{\tilde t_2},m_{\tilde g},m_t)\right]\Biggr\}\ ,
\label{2lep}
\end{eqnarray}
where the arguments 
$h_1, h_2$ are the neutral scalar components of the Higgs fields
$H_1$ and $H_2$. All the masses and mixing angles
in Eq.~(\ref{2lep}) should
be understood as $h_1, h_2$-dependent quantities, 
for example, the top quark mass
$m_t={1\over\sqrt{2}}h_t h_2$.
We use short-hand notations 
$c_t=\cos\theta_{\tilt}$, 
$s_t=\sin\theta_{\tilt}$, where $\theta_{\tilt}$ is the
top-squark mixing angle.
The minimally subtracted\footnote{Subtraction of
one-loop sub-divergences can be treated following
the first reference of \cite{FJJ}. 
In obtaining Eq.~(\ref{2lep}), we have also
added freely one-loop diagrams depending only on $m_{\tilde g}$,
as they do not affect the Higgs boson mass calculation in the 
effective potential approach.} two-loop scalar functions $I$ and $J$ 
are \cite{FJJ}:
\begin{eqnarray}
&&I(m_1,m_2,m_3)\ =\ -{1\over 2}\biggl[(-m_1^2+m_2^2+m_3^2)
\ln{m^2_2\over Q^2}\ln{m^2_3\over Q^2}\nonumber\\
&& ~~+
(m_1^2-m_2^2+m_3^2) \ln{m^2_1\over Q^2}\ln{m^2_3\over Q^2}
+(m_1^2+m_2^2-m_3^2) \ln{m^2_1\over Q^2}\ln{m^2_2\over Q^2}
\nonumber\\
&&~~-4(m^2_1\ln {m^2_1\over Q^2}+m^2_2\ln {m^2_2\over Q^2}
+m^2_3\ln {m^2_3\over Q^2})+\xi(m_1,m_2,m_3)+5(m^2_1+m^2_2+m^2_3)
\biggr]\ ,\vspace{0.1cm}\\
&&J(m_1,m_2)\ =\ m^2_1 m^2_2\biggl[1-\ln{m_1^2\over Q^2}
-\ln{m_2^2\over Q^2}+\ln{m_1^2\over Q^2}\ln{m_2^2\over Q^2}\biggr]\
,\vspace{0.1cm}
\end{eqnarray}
where $Q$ is the renormalization scale.
The function $\xi(m_1,m_2,m_3)$ in the expression of $I(m_1,m_2,m_3)$ 
has been calculated in \cite{FJJ} by a differential equation method
and in \cite{DT} by a Mellin-Barnes integral representation method. 
Its final form can be expressed in terms of 
Lobachevsky's functions or Clausen's integral functions and their
analytical continuations.

It is important to check that the effective potential 
$V(h_1,h_2) = V_{0}(h_1,h_2)+V_{1}(h_1,h_2)
+V_{2}(h_1,h_2)$ is invariant under changes of the renormalization
scale up to higher order terms of ${\cal O}(\alpha_t^2)$.
To prove this, we write the tree-level and one-loop potential as
follows (neglecting contributions from the gauge boson,
Higgs/Goldstone boson and neutralino/chargino sectors):
\begin{equation}
V_{0+1}(h_1,h_2)\ =\ 
{1\over2}(m^2_{H_1}+\mu^2)h^2_1+{1\over2}(m^2_{H_2}+\mu^2)h^2_2
+B_\mu h_1 h_2 + {3\over\olf}\biggl[G(m_{\tilt_1})+G(m_{\tilt_2})
-2G(m_t)\biggr]\ ,
\end{equation}
where $m_{H_1}, m_{H_2}$ and $B_\mu$ are the soft-breaking
Higgs sector mass parameters, $\mu$ the supersymmetric Higgs-boson
mass parameter, and
\begin{equation}
G(m)\ =\ {m^4\over2}\biggl(\ln{m^2\over Q^2} - {3\over 2}\biggr)\ .
\end{equation}
It is then straightforward to show that
\begin{equation}
{\cal D}^{(2)}V_{0}\ =\ -{\partial V_{2}\over\partial\ln Q^2}
-{\cal D}^{(1)}V_{1}\ =\ 
{8g_3^2h_t^2h_2^2\over(16\pi^2)^2}
\biggl(\MQ^2+\MU^2+2M_3^2+\tAt^2
-2M_3\tAt\biggr)\ ,
\label{eq:rgv}
\end{equation}
modulo terms independent of the Higgs field $h_2$, and where
$\tAt= A_t+\mu\cot\beta$, with $A_t$ a trilinear 
coupling (with dimensions of mass) appearing in the following term of the
soft-breaking
Lagrangian: $h_t A_t H_2 \widetilde{Q} \widetilde{U}$.
$\MQ$ and $\MU$ are soft-breaking masses 
for the left- and right-handed top squarks, $\widetilde{Q}$ 
and $\widetilde{U}$, respectively, and $M_3$ is the gluino soft mass.
[It is a nontrivial check that all $\ln Q^2$ terms cancel with each
other in Eq.~(\ref{eq:rgv})]. In the above equation
${\cal D}^{(1)}$ and ${\cal D}^{(2)}$ stand for one- and two-loop
RG variations of the parameters, and we have used the following equations
\begin{eqnarray}
&&{\partial I(m_1,m_2,m_3)\over\partial\ln Q^2}
\ =\ m^2_1\ln {m^2_1\over Q^2}+m^2_2\ln {m^2_2\over Q^2}
+m^2_3\ln {m^2_3\over Q^2} -2 (m^2_1+m^2_2+m^2_3)\ ,\\
&&{\partial J(m_1,m_2)\over\partial\ln Q^2}
\ =\ m^2_1m^2_2\biggl(2-\ln{m^2_1\over Q^2} -\ln{m^2_2\over Q^2}\biggr)\ ,
\end{eqnarray}
and the MSSM RG equations from Ref.~\cite{martin},
\begin{eqnarray}
{\partial m_{H_2}^2\over\partial\ln Q^2} &=&~
{3h_t^2\over16\pi^2}(m^2_{H_2}+\MQ^2+\MU^2+A_t^2)\nonumber\\
&&+{16g_3^2h_t^2\over(16\pi^2)^2}
(m^2_{H_2}+\MQ^2+\MU^2+A_t^2+2M_3^2-2M_3A_t)\ ,\\
{\partial\ln\mu\over\partial\ln Q^2} &=& 
- {\partial\ln h_2\over\partial\ln Q^2}\ =\
{3h_t^2\over32\pi^2}+{8g_3^2h_t^2\over(16\pi^2)^2}\ ,\\
{\partial B_\mu\over\partial\ln Q^2} &=&
{3h_t^2\over16\pi^2}\left({B_\mu\over2}+A_t\mu\right)
+{16g_3^2h_t^2\over(16\pi^2)^2}
\left({B_\mu\over2}+A_t\mu-M_3\mu\right)\ .
\end{eqnarray}
We will see that the RG does not only provide an important 
check for the Higgs boson mass correction formulae in the later sections,
but also allows us to present those formulae in a more physically
appealing form$-$the RG improved form.

Using the effective potential in Eq.~(\ref{2lep}),
two-loop radiative corrections to 
$m^2_{h^0}$ [to the order of ${\cal O}(\alpha_s\alpha_t)$]
have been calculated numerically in \cite{RenJie}. In the following,
we shall derive an approximation formula valid for $\MQ(=\MU)$ and
$m_{A^0}\gg M_Z$, where $m_{A^0}$ is the mass of the pseudoscalar Higgs boson
$A^0$.

\section{Analytical expression for $\Delta m_{h^0}^2$ from the effective
potential}

To simplify our analytical expression,
we assume the squark soft masses satisfy $\MQ=\MU\equiv M_S$, 
where $M_S$ is the SUSY scale.
The two eigenvalues and mixing angle of 
top-squark squared-mass matrix are 
\begin{equation}
m^2_{{\tilt}_1}\ =\ \mtilt^2+m_t\tAt\ ,\qquad
m^2_{{\tilt}_2}\ =\ \mtilt^2-m_t\tAt\ ,\qquad
s_t\ =\ c_t\ =\ {1\over\sqrt{2}}\ ,
\end{equation}
where the average top-squark mass $\mtilt^2=M_S^2+m_t^2$.
We shall further assume the gluino mass $m_{\tilde g}=M_S$.

Under these simplifications, the effective potential to 
${\cal O}(\alpha_s\alpha_t)$ two-loop order is 
(neglecting one-loop sub-dominant terms)
\begin{eqnarray}
V(h_1,~h_2) &=& V_{0}(h_1,~h_2)
+ {3\over\olf}\biggl[2G(\mtilt)-2G(m_t)
+m^2_t\tAt^2\ln {\mtilt^2\over Q^2} 
- {m^4_t\tAt^4\over 12 \mtilt^4}
\biggr]\nonumber\\
&+&{8g_3^2\over(\olf)^2}\Biggl\{J(m_t,m_t)-2m^2_t I(m_t,m_t,0) 
+ J(\mtilt,\mtilt) +2\mtilt^2 I(\mtilt,\mtilt,0)\nonumber\\
&&\qquad\qquad + 2J(M_S,m_t)-2J(M_S,\mtilt)-2J(m_t,\mtilt)\nonumber\\
&+&m^2_t\biggl[2\tAt\mtilt 
(1-\ln{\mtilt^2\over Q^2})^2
+\tAt^2(1-\ln^2{\mtilt^2\over Q^2})\biggr]\nonumber\\ 
&+&m_t^4\biggl[-{\tAt\over\mtilt}\biggl((1-\lnttq)^2+2\lntt\biggr)
+{\tAt^2\over\mtilt^2}+{\tAt^3\over3\mtilt^3}\biggl(1-2\lnttq\biggr)
-{\tAt^4\over12\mtilt^4}\biggr]\Biggr\},
\label{pot}
\end{eqnarray}
where we have separated terms depending on $\tAt$ (to make 
the top-squark mixing effects transparent) and expanded the
effective potential in powers of $m_t/\mtilt$ (and $m_t X_t/\mtilt^2$) with
higher order terms
in $m_t/\mtilt$ neglected.
It is now straightforward to find the corrections to the four entries in the 
Higgs boson squared-mass matrix\footnote{The effective potential approach 
evaluates these corrections at the zero external momentum limit. This is
consistent with our other approximations since the 
tree-level light Higgs boson mass is related to the 
electroweak gauge couplings and can be neglected.} 
\begin{eqnarray}
\Delta{\cal M}^2_{11} &=& \left. h_1{\partial\over\partial h_1}
\left({1\over h_1}{\partial V(h_1,~h_2)\over\partial h_1}\right)
\right|_{h_1=v_1, h_2=v_2}
\nonumber\\
&=& {\alpha_t\over4\pi}\biggl\{3\tan\beta
\biggl[-\mu A_t\ln{\mtilt^2\over Q^2}+{m_t^2\mu A_t\tAt^2\over 6\mtilt^4}
\biggr]-{m_t^2\mu^2\tAt^2\over\mtilt^4}\biggr\}\nonumber\\
&&+{\alpha_s\alpha_t\over2\pi^2}
\Biggl\{-\mu\tan\beta\biggl[\mtilt
\biggl(1-\ln{\mtilt^2\over Q^2}\biggr)^2
+A_t\biggl(1-\ln^2{\mtilt^2\over Q^2}\biggr)\biggr]\nonumber\\
&&+{m^2_t\mu\over2\mtilt}\tan\beta
\biggl[(1-\lnttq)^2+2\lntt\biggr]
-{m^2_t\mu A_t\over\mtilt^2}\tan\beta\nonumber\\
&&
+{m^2_t\tAt\over\mtilt^3}(\mu^2-{\mu\tAt\over2}\tan\beta)
\biggl[1-2\lnttq\biggr]
+{m^2_t\mu\tAt^2\over2\mtilt^4}
\biggl[-\mu+{\tAt\over3}\tan\beta\biggr]\Biggr\}\ ,\\
\Delta{\cal M}^2_{12} &=& \Delta{\cal M}^2_{21}\ =\
\left. {\partial^2 V(h_1,~h_2)\over\partial h_1\partial
h_2}\right|_{h_1=v_1, h_2=v_2} \nonumber\\
&=&
{3\alpha_t\over4\pi}
\biggl[\mu A_t\ln{\mtilt^2\over Q^2}
+{2m_t^2\mu\tAt\over\mtilt^2}
-{m_t^2\mu A_t\tAt^2\over2\mtilt^4}\biggr]\nonumber\\
&&+{\alpha_s\alpha_t\over2\pi^2}
\Biggl\{\mu\biggl[\mtilt 
\biggl(1-\ln{\mtilt^2\over Q^2}\biggr)^2 +A_t
\biggl(1-\ln^2{\mtilt^2\over Q^2}\biggr)\biggr]
\nonumber\\
&&+{m^2_t\mu\over\mtilt}
\biggl[-{5\over2}+5\ln{\mtilt^2\over Q^2}
-{1\over2}\ln^2{\mtilt^2\over Q^2}-3\lntt\biggr]
+{m_t^2\mu\over\mtilt^2}\biggl[A_t+2\tAt-4\tAt\lnttq\biggr]
\nonumber\\
&&+{m^2_t\tAt\over\mtilt^3}(\mu A_t+{\mu\tAt\over2})
\biggl[1-2\lnttq\biggr]
-{m^2_t\mu A_t\tAt^2\over2\mtilt^4}\Biggr\}\ ,\\
\Delta{\cal M}^2_{22} &=& \left. h_2{\partial\over\partial h_2}
\left({1\over h_2}{\partial V(h_1,~h_2)\over\partial h_2}\right)
\right|_{h_1=v_1, h_2=v_2}
\nonumber\\
&=&{3\alpha_t\over\pi} m_t^2\ln{\mtilt^2\over m^2_t}
+{\alpha_t\over4\pi}\Biggl\{3\cot\beta\biggl[-\mu A_t\ln{\mtilt^2\over Q^2}
+{m_t^2\mu A_t\tAt^2\over 6\mtilt^4}\biggr]
+\biggl[{12A_t\tAt\over\mtilt^2}
-{A_t^2\tAt^2\over\mtilt^4}\biggr]\Biggr\}\nonumber\\
&&+{2\alpha_s\alpha_t\over\pi^2}m_t^2
\biggl[\ln^2{\mtilt^2\over m^2_t}-2\ln^2{\mtilt^2\over Q^2}
+2\ln^2 {m^2_t\over Q^2}
+\ln{\mtilt^2\over m^2_t} -1\biggr]\nonumber\\
&&+{\alpha_s\alpha_t\over2\pi^2}
\Biggl\{-\mu\cot\beta\biggl[\mtilt
\biggl(1-\ln{\mtilt^2\over Q^2}\biggr)^2
+A_t(1-\ln^2{\mtilt^2\over Q^2})\biggr]\nonumber\\
&&+{m^2_t\over\mtilt}
\biggl[{\mu\over2}\cot\beta
\biggl((1-\lnttq)^2+2\lntt\biggr)+2(A_t+\tAt)
\biggl(4\lnttq-2\lntt\biggr)-4 A_t\biggr]\nonumber\\
&&+{m^2_tA_t\over\mtilt^2}\biggl[4\tAt-\mu\cot\beta
-8\tAt\lnttq\biggr]
+{m^2_t\tAt\over\mtilt^3}\biggl[A^2_t-{\mu\tAt\over2}\cot\beta
+{\tAt^2\over3}\biggr]\biggl[1-2\lnttq\biggr]\nonumber\\
&&+{m^2_tA_t\tAt^2\over2\mtilt^4}
\biggl[-A_t+{\tAt\over3}\biggr]\Biggr\}\ ,
\end{eqnarray}
where $\alpha_s=g^2_3/4\pi$ and $\alpha_t=h^2_t/4\pi$.
Again we have neglected terms in higher orders of $m_t/\mtilt$.

The Higgs boson mass to the two-loop order can be obtained by
diagonalizing the CP-even Higgs boson
squared-mass matrix ${\cal M}^2$ by including
these corrections
\begin{equation}
{\cal M}^2\ =\ \left(
\begin{array}{cc}
M^2_Z \cos^2\beta + m^2_{A^0} \sin^2\beta + \Delta{\cal M}^2_{11} &
-(M^2_Z + m^2_{A^0}) \sin\beta \cos\beta + \Delta{\cal M}^2_{12} \\
-(M^2_Z + m^2_{A^0}) \sin\beta \cos\beta + \Delta{\cal M}^2_{21} &
M^2_Z \sin^2\beta + m^2_{A^0} \cos^2\beta + \Delta{\cal M}^2_{22}
\end{array}\right)\ .
\end{equation}
This means that the mixing angles at two-loop
order are different from those at tree-level. Nevertheless, 
this difference is small to a good approximation and we can therefore use
the tree-level mixing angle $\alpha$ to compute the Higgs boson mass.
We further simplify the mass correction formula by approximating
$\cos\alpha=\sin\beta$ and $\sin\alpha=-\cos\beta$, which are valid in the
limit $m_{A^0}\gg M_Z$. The final analytical expression for the
two-loop order correction to the lightest CP-even Higgs boson mass is
\begin{eqnarray}
\Delta m^2_{h^0} &=& 
{3m^4_t\over2\pi^2 v^2}\biggl[\ln{\mtilt^2\over m_t^2}
+{\tAt^2\over\mtilt^2}
-{\tAt^4\over12\mtilt^4}\biggr]
+{\alpha_sm^4_t\over\pi^3 v^2}
\Biggl\{\ln^2{\mtilt^2\over m^2_t}-2\ln^2{\mtilt^2\over Q^2}
+2\ln^2 {m^2_t\over Q^2}
+\ln{\mtilt^2\over m^2_t} -1\nonumber\\
&&+{\tAt\over\mtilt}\biggl[-1+2\lnttq+2\lntq\biggr]
+\biggl({\tAt^2\over\mtilt^2}+{\tAt^3\over3\mtilt^3}\biggr)
\biggl[1-2\lnttq\biggr]
-{\tAt^4\over12\mtilt^4}\Biggr\}\ .
\label{eq:dmdr}
\end{eqnarray}
We note that all parameters in Eq.~(\ref{eq:dmdr}) are running
parameters evaluated in the 
$\dr$-scheme and satisfy MSSM RG equations. It is also easy to check that
$\Delta m_{h^0}^2$ is indeed
two-loop RG invariant up to terms of order ${\cal O}(\alpha_t^2)$ 
by using the following equations
\begin{equation}
{\partial\ln m^2_t\over\partial\ln Q^2}\ =\ 
{\partial\ln \mtilt^2\over\partial\ln Q^2}\ =\ -{4\alpha_s\over3\pi}\ ,
\qquad
{\partial \tAt\over\partial\ln Q^2}\ =\ {4\alpha_s\over3\pi} M_3\ ,
\label{eq:rge}
\end{equation}
with the gluino soft mass $M_3=m_{\tilde g}=M_S$ in our approximation.

At first sight, Eq.~(\ref{eq:dmdr}) seems quite different from
those formulae obtained by the RG improved one-loop effective potential
approach \cite{CEQWHHH}
and by the two-loop diagrammatic approach \cite{HHWan}. The difference
can be settled by observing that the parameters in Eq.~(\ref{eq:dmdr})
are MSSM running parameters, while the top quark mass in \cite{CEQWHHH}
is a SM running mass and $\tAt$ and $\mtilt$ in \cite{HHWan} are
on-shell (OS) parameters. 
Our one-loop parameters receive radiative corrections from the SM 
as well as SUSY particles, so they are different from the parameters
in other approaches where the SUSY particles are explicitly decoupled.
Furthermore, part of the difference arises from converting parameters in
the $\dr$-scheme to the $\ms$-scheme.

To demonstrate this point, we first compare our result in Eq.~(\ref{eq:dmdr})
with that of Ref. \cite{HHWan}. 
We use the following relations of $\dr$ top quark mass ($m_t$) and 
top-squark running mass ($\mtilt$) (in the MSSM) 
and the $\ms$ running top quark mass ($\bmt$) 
and OS top-squark mass ($\hmst$) (in the SM), 
which for example can be
deduced from Refs. \cite{PBMZ, Donini}
\begin{eqnarray}
m_t(Q) &=& 
\bmt(Q)\biggl[1 
- {\alpha_s\over3\pi}\biggl(1+\lnttq-{\tAt\over\mtilt}\biggr)\biggr]\ ,
\label{eq:mtconv}\\
\mtilt(Q) &=& \hmst\biggl[1-
{4\alpha_s\over3\pi} + {2\alpha_s\over3\pi} \ln{\mtilt^2\over Q^2}\biggr]\ ,
\label{eq:mstconv}
\end{eqnarray}
up to higher order terms in $m_t/\mtilt$.\footnote{Parameters inside 
the brackets can be running or OS ones, as the difference is of higher
order.}
All terms in Eqs.~(\ref{eq:mtconv}) and (\ref{eq:mstconv}) 
come from the decoupling top squarks and gluinos, except for
the finite term $-\alpha_s/3\pi$ in Eq.~(\ref{eq:mtconv}), which  
is a $\dr$-$\ms$-scheme conversion factor. Similarly, one
can relate $\tAt$ between different approaches by considering
the top-squark/gluino contributions to the self energy 
$\Pi_{\tilt_L\tilt_R}$ \cite{PBMZ}, which gives
\begin{eqnarray}
\tAt(Q)\ &=& \ 
\htAt + {1\over m_t}\biggl[{\rm Re}\Pi_{\tilt_L\tilt_R}(p^2=\mtilt^2)
- \tAt{\rm Re}\Sigma_t(p^2=m_t^2)\biggr]\nonumber\\
&=& \htAt
+ {\alpha_s\over3\pi}\biggl[\mtilt\left(8-4\lnttq\right)
+\tAt\left(5+3\lntt\right)-{\tAt^2\over\mtilt}\biggr]\ ,
\label{eq:atconv}
\end{eqnarray}
where $\htAt$ is the OS top-squark mixing parameter.
The external momentum of $\Pi_{\tilt_L\tilt_R}$ has been set to
the average top-squark mass $\mtilt$.\footnote{Choosing $p^2=\mtilt^2\pm 
m_tX_t$ changes the result by terms of higher order in 
$m_t/\mtilt$.} The top-quark self energy 
$\Sigma_t$ is evaluated in the MSSM and contains QCD corrections
from top-squark/gluino and top/gluon loops 
(it can be deduced from Refs.~\cite{PBMZ, Donini}),
\begin{equation}
{\rm Re}\Sigma_t(p^2=m_t^2)\ =\
-{\alpha_s\over3\pi}m_t\biggl[5-3\lntq+\lnttq
-{\tAt\over\mtilt}\biggr]\ .
\end{equation} 
We note that in the effective 
SM the average top-squark mass $\hmst$ and the parameter
$\htAt$ are frozen and equivalent to their physical `pole' values
since all SUSY particles have been decoupled.

Substituting Eqs.~(\ref{eq:mtconv}), (\ref{eq:mstconv}) and
(\ref{eq:atconv}) into Eq.~(\ref{eq:dmdr}), 
we obtain the mass correction formula 
\begin{eqnarray}
\Delta m_{h^0}^2 &=& 
{3\bmt^4\over2\pi^2 v^2}
\biggl[\ln{\hmst^2\over \bmt^2}
+{\htAt^2\over\hmst^2}
-\frac{1}{12}{\htAt^4\over\hmst^4}\biggr]
+{\alpha_s m^4_t\over\pi^3 v^2}
\Biggl\{-4-3\ln^2{\mtilt^2\over Q^2}+3\ln^2 {m_t^2\over Q^2}
+3\ln{\mtilt^2\over Q^2}-\ln{\mtilt^2\over m^2_t}\nonumber\\
&&+{6\tAt\over\mtilt}
+{\tAt^2\over\mtilt^2}\biggl[8-3\lntq-3\lnttq\biggr]
+{\tAt^4\over\mtilt^4}\biggl[-{17\over12}+{1\over2}\lntq\biggr]\Biggr\}\ ,
\label{eq:dmsm}
\end{eqnarray}
where we have explicitly indicated the definition used for different
parameters in the one-loop corrections; different definitions
for the parameters in the two-loop corrections 
would give differences of higher order.
This equation is identical
to the mass correction formula in \cite{HHWan} for $Q=m_t$.
Comparison of our formula with the
one obtained using the RG-improved one-loop potential \cite{CEQWHHH}
is left for the end of next section.

\section{RG-improved Higgs boson mass}

The final expression for $m_{h^0}^2$ in the last section was very
convenient to make contact with the diagrammatic results of \cite{HHWan}.
However, a simpler and more transparent expression can be obtained by
transforming to the RG language. Furthermore, the improvement of the
formula goes beyond purely aesthetic reasons, as it resums higher order
corrections
as well. 

The idea is to let all parameters in the formula for $m_{h^0}^2$ be
running parameters and to choose the scale at which they are evaluated in
such a way that higher order logarithmic corrections are automatically 
taken care of. Moreover, the scale at which each parameter has to be
evaluated to achieve this, is susceptible of physical interpretation.

Starting from Eq.~(\ref{eq:dmdr}), we first use Eq. (\ref{eq:mtconv}) to
translate the MSSM top-quark running mass $m_t(Q)$ into that of the SM, 
$\bmt(Q)$. We then
use the solutions to the SM top quark mass RG equation  
and the MSSM RG equations of $\mtilt$ and $\tAt$ in Eq. (\ref{eq:rge}),
\bear
\label{mtoprun}
\bmt^2(Q)&=&\bmt^2(Q')
\left[1+\frac{2\alpha_s}{\pi}\ln\frac{{Q'}^2}{Q^2}\right]\ ,\\
\label{mstrun}
\mtilt^2(Q)
&=&\mtilt^2(Q')\left[1+\frac{4\alpha_s}{3\pi}\ln\frac{{Q'}^2}{Q^2}\right]\ ,\\
\label{matrun}
\tAt(Q)
&=&\tAt(Q') - \frac{4\alpha_s}{3\pi}M_3\ln\frac{{Q'}^2}{Q^2}\ ,
\eear
to relate their values at two different renormalization scales, $Q$ and
$Q'$. 
Our final formula is
\bear
\label{mhrg}
\Delta m_{h^0}^2&=&\frac{3}{2\pi^2v^2}\left\{\bmt^4(Q_t)
\ln\frac{m_{\tilde{t}}^2(Q_{\tilde{t}})}{\bmt^2(Q'_t)}
+\bmt^4(Q_{\rm th})\left[
\frac{\tAt^2(Q_{\rm th})}{m_{\tilde{t}}^2(Q_{\rm th})}-
\frac{\tAt^4(Q_{\rm th})}{12
m_{\tilde{t}}^4(Q_{\rm th})}\right]\right\}\nonumber\\
&+&\frac{\alpha_s m_t^4}{\pi^3v^2}\left[
-\frac{2\tAt}{m_{\tilde{t}}}
-\frac{\tAt^2}{m_{\tilde{t}}^2}
+\frac{7}{3}\frac{\tAt^3}{m_{\tilde{t}}^3}
+\frac{1}{12}\frac{\tAt^4}{m_{\tilde{t}}^4}
-\frac{1}{6}\frac{\tAt^5}{m_{\tilde{t}}^5}
\right],
\eear
with $Q_t^2=m_t m_{\tilde{t}}$, $Q_{\tilde{t}}^2 m_t={Q'}^3_t$
(satisfied by $Q_{\tilde{t}}={Q'}_t=m_t$
or $Q_{\tilde{t}}=m_{\tilde{t}}$, ${Q'}^3_t=m_t m_{\tilde{t}}^2$) and
$Q_{\rm th}=m_{\tilde{t}}$.
Note how nicely everything falls into place. All higher order logarithms are
reabsorbed into the one-loop RG-improved term and the only two-loop pieces
remaining are the finite $\tAt$-dependent terms.

The logarithmic term in Eq. (\ref{mhrg}) can be interpreted as the result of
integrating the
RG equation for the quartic Higgs coupling between the high energy scale
$m_{\tilde{t}}$ and the top-quark mass scale $m_t$ in the SM, which
is the effective theory below $m_{\tilde{t}}$. The coupling in front of
this logarithmic term has to be evaluated 
at the intermediate scale $Q_t=\sqrt{m_t m_{\tilde{t}}}$, 
choice which automatically takes into account higher
order effects. Having computed the two-loop results, we can also say
something on
the scale at which the masses entering the logarithm should be evaluated,
although not in an unambiguous way. If we insist in evaluating them at the
same scale, that scale turns out to be $Q_{\tilde{t}}={Q'}_t=m_t$. If, on
the other
hand, we prefer to keep $Q_{\tilde{t}}=m_{\tilde{t}}$, then ${Q'}_t=(m_t
m_{\tilde{t}}^2)^{1/3}$. The finite non-logarithmic `one-loop' term is
interpreted as a threshold correction for the quartic Higgs coupling at the
scale $m_{\tilde{t}}$, at which the SM is matched to the MSSM. 
In accordance to this, all the parameters in this term are evaluated at the
threshold scale $Q_{\rm th}=m_{\tilde{t}}$. Then, the remaining two-loop finite
terms give a two-loop contribution to this 
threshold correction.\footnote{Actually, 
an alternative way of computing this two-loop threshold
correction is the following. Start with the expression for the effective
potential in Eq. (\ref{pot}) and make use of Eq. (\ref{eq:mtconv}) 
to trade between
the running masses in MSSM and SM. Expand then the potential in powers of
$m_t$. In that expansion, the finite term of order $m_t^4$ gives the
correction to the Higgs quartic coupling, in accordance with our final
result Eq. (\ref{mhrg}).}
Eq. (\ref{mhrg}) can be slightly modified to automatically
include the logarithmic
terms of order ${\cal O}(\alpha_t^2)$ (and higher) if we refine
Eqs.~(\ref{eq:mtconv}), (\ref{mtoprun}), (\ref{mstrun}) and 
(\ref{matrun}) to take into
account ${\cal O}(\alpha_t)$ corrections.

\begin{figure}[tbh]
\postscript{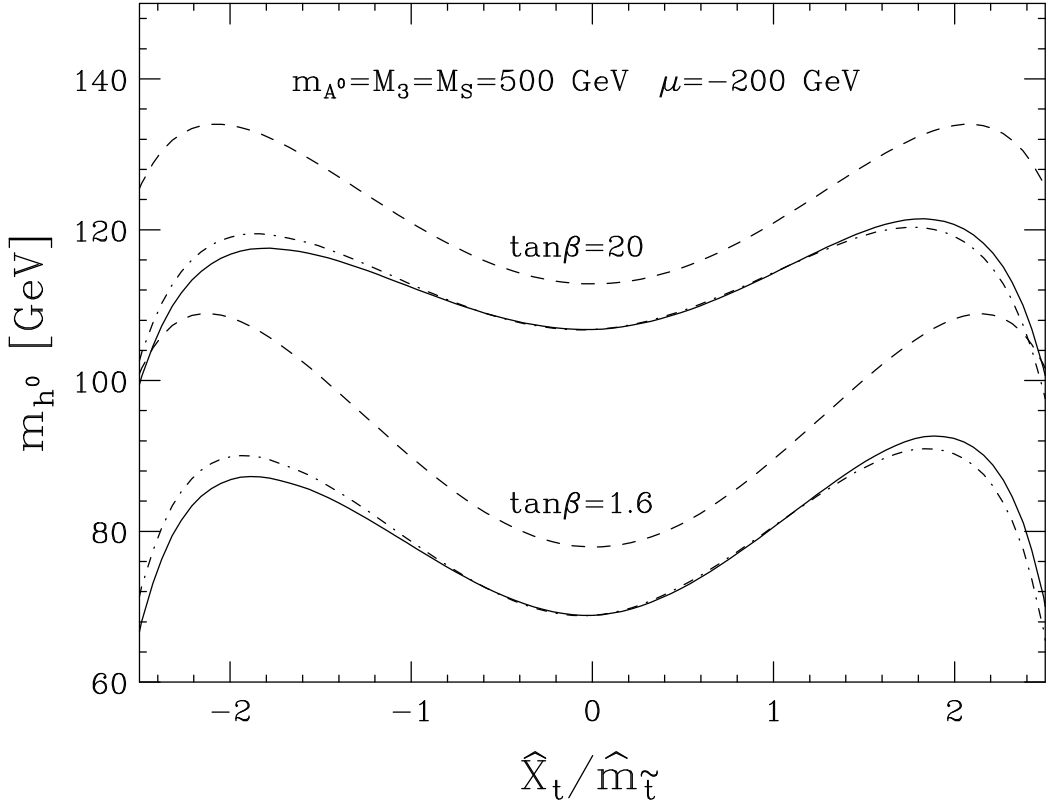}{0.8}
\caption[]{Higgs boson mass $m_{h^0}$ versus $\htAt/\hmst$. 
The solid lines and dotdashes correspond to the two-loop Higgs
boson masses calculated from the RG-improved mass formula Eq. (\ref{mhrg}),
with and without including the two-loop finite threshold corrections.
One-loop masses are shown in dashes for reference.} 
\label{fig1}
\end{figure}

We show in Fig.~\ref{fig1} the numerical result from the RG-improved
Higgs boson mass formula Eq. (\ref{mhrg}). 
The two-loop finite threshold corrections are generally small for small and
moderate mixing parameter $\htAt/\hmst$. For large mixing 
$\htAt/\hmst\gsim 2$, however, the two-loop threshold corrections can 
contribute corrections of the size of $\sim$ 3 GeV.

In terms of the parameter
$\tAt(Q_{\rm th})/m_{\tilde{t}}(Q_{\rm th})$, we can obtain
the refined condition for maximal mixing as
\be
\left[\frac{\tAt(Q_{\rm th})}{m_{\tilde{t}}(Q_{\rm th})}\right]_{\rm max}=
\pm\sqrt{6}+\frac{5\alpha_s}{3\pi}\ ,
\ee
which corresponds to a $2\%$ shift to the position of maximal mixing.
This condition can be rewritten in terms of OS quantities as
\be
\left[\frac{\htAt}{\hmst}\right]_{\rm
max}=
\pm\sqrt{6}\left[1-\frac{\alpha_s}{\pi}
\left(3+\ln\frac{m^2_{\tilde{t}}}{m_t^2}\right)\right]+\frac{\alpha_s}{\pi}.
\ee

We are now ready to compare with the formulae presented in \cite{CEQWHHH}
using the RG-improved one-loop potential (some two-loop input was used
in the last paper in \cite{CEQWHHH}).
For zero squark-mixing, the leading and next-to-leading logarithmic
terms in Eq.~(\ref{eq:dmsm}) 
agree with \cite{CEQWHHH} but those papers also gave the 
RG-improved one-loop threshold correction for non-zero top-squark
mixing case, which is [omitting the ${\cal O}(\alpha_t^2)$ terms]
\begin{equation}
{3\bmt^4(m_t)\over2\pi^2 v^2} 
\biggl[{\tAt^2\over\mtilt^2}
-{\tAt^4\over12\mtilt^4}\biggr]\biggl[1-
{4\alpha_s\over\pi}\lntt\biggr]\ .
\label{RGimprov}
\end{equation}
In this formula the $\tAt$ parameter is implicitly evaluated in the MSSM
at the threshold scale, exactly as in our formula Eq. (\ref{mhrg}), and
agreement between the two is found after expressing  $\bmt(m_t)$
in terms of $\bmt(Q_{\rm th})$.
However, the two-loop finite terms included in Eq. (\ref{mhrg}) can not be
reproduced
by the RG-improved one-loop effective potential approach. 

To summarize, our analytical expression for $m^2_{h^0}$ agrees with
those obtained by explicit two-loop calculation \cite{HHWan} and
RG-improved one-loop effective potential approach \cite{CEQWHHH}
where they overlap. We note that parameters in 
different models ({\it i.e.}, MSSM and SM) and renormalization
schemes have been used in previous studies, our discussions
should have resolved any possible confusion.

\section{Conclusions and outlook}

We have used the knowledge of the MSSM effective potential up to two-loop
${\cal O}(\alpha_s\alpha_t)$ to extract a simple analytical approximation
to the mass $m_{h^0}$ of the lightest CP-even
Higgs boson in the simple case of a single
supersymmetric threshold significantly higher than $M_Z$. We have
considered the case of arbitrary $\tan\beta$ and non-zero mixing 
in the top-squark sector. We have derived a RG-improved formula to
resum large logarithmic corrections, achieving a particularly simple and
illuminating final result.

Our results agree with previous analyses based on the RG-improved one-loop
effective potential \cite{CEQWHHH} and
diagrammatic calculation \cite{HHW} to the order at which such studies
were
performed. By doing this we clarify the relation between different
approaches and identify the two-loop origin of the discrepancies
between \cite{CEQWHHH} and \cite{HHW} for large values of the top-squark
mixing. This difference can be attributed to a two-loop threshold
correction, most easily calculable in the effective potential approach,
as we have shown. We emphasize that in applying two-loop mass
correction formulae, one should be particularly careful about the
parameters entering the one-loop formulae;
we have done this by differentiating running and OS parameters
in our mass correction formulae, Eqs. (\ref{eq:dmdr}), (\ref{eq:dmsm}) and
(\ref{mhrg}).

The combined use of the three different techniques (effective potential,
diagrammatic calculation and RG-resummation) is very powerful and should 
be applied to compute the still missing ${\cal O}(\alpha_t^2)$ corrections
to $m_{h^0}^2$, which can be similar in magnitude to those analyzed in
this paper. Phenomenological analyses done with the expressions 
currently given in the literature are accurate only up to the inclusion of
these
${\cal O}(\alpha_t^2)$ corrections, which one could expect to give a
shift in $m_{h^0}$ not greater than 5 GeV, but quite interestingly,
we expect that this shift goes in a direction opposite to that of the 
two-loop ${\cal O}(\alpha_s\alpha_t)$
corrections. The leading and next-to-leading logarithmic corrections
may be obtained by a RG resummation approach, but the finite
non-logarithmic corrections can only be extracted by a direct 
two-loop calculation to the order ${\cal O}(\alpha_t^2)$, {\it e.g.},
from effective potential or from explicit diagrammatic calculation.

\section*{Note Added}
A comparison between the diagrammatic and RG results for $m_{h^0}^2$ is
mentioned in M.~Carena, S.~Heinemeyer, C.E.M.~Wagner and G.~Weiglein,
[hep-ph/9912223], which appeared at the time of submitting this paper.

\vspace{.7cm}
{\noindent\Large\bf Acknowledgments}
\vspace{.7cm}

J.R.E. would like to thank H.E. Haber and A.H. Hoang for discussions. 
R.-J.Z. would like to thank T. Falk, H.E. Haber, T. Han, S. Heinemeyer, T.
Plehn and C. Wagner for conversations. The research of R.-J.Z. was
supported in part by a DOE grant No. DE-FG02-95ER40896 and in part by the
Wisconsin Alumni Research Foundation.


\begin{thebibliography}{99}
%
\bibitem{lh} P.~Langacker and H.~A.~Weldon, \PRL{52}{84}{1377};
H.~A.~Weldon, \PLB{146}{84}{59};
D.~Comelli and J.~R.~Espinosa, \PLB{388}{96}{793}.
%
\bibitem{eg} J.~R.~Espinosa and J.~F.~Gunion, \PRL{82}{99}{1084}.
%
\bibitem{LEP2} LEP Experiments Committee Meeting, Nov. 9th, 1999,\\
{\tt
http://delphiwww.cern.ch/$\sim$offline/physics$\underline{~}$links/lepc.html}.
%
\bibitem{rad} S.~P.~Li and M.~Sher, \PLB{140}{84}{339};
M.~S.~Berger, \PRD{41}{90}{225};
Y.~Okada, M.~Yamaguchi, and T.~Yanagida, \PTP{85}{91}{1}; \PLB{262}{91}{54};
J.~Ellis, G.~Ridolfi and F.~Zwirner, \PLB{257}{91}{83}; \PLB{262}{91}{477};
H.~E.~Haber and R.~Hempfling, \PRL{66}{91}{1815};
R.~Barbieri, M.~Frigeni and M.~Caravaglios, \PLB{258}{91}{167};
J.~R.~Espinosa and M.~Quir\'{o}s, \PLB{266}{91}{389};
J.~L.~Lopez and D.~V.~Nanopoulos, \PLB{266}{91}{397};
D.~M.~Pierce, A.~Papadopoulos and S.~B.~Johnson, \PRL{68}{92}{3678};
A.~Brignole, \PLB{281}{92}{284};
M.~Drees and M.~M.~Nojiri, \PRD{45}{92}{2482}; 
J.~Kodaira, Y.~Yasui and K.~Sasaki, \PRD{50}{94}{7035};
P.~H.~Chankowski, S.~Pokorski and J.~Rosiek,
\NPB{423}{94}{437}; A.~V.~Gladyshev and D.~I.~Kazakov, \MPLA{10}{95}{3129};
A.~Dabelstein, \ZPC{67}{95}{495};
J.~A.~Casas, J.~R.~Espinosa, M.~Quir\'os and A.~Riotto, 
\NPB{436}{95}{3}; A.~V.~Gladyshev, D.~I.~Kazakov, 
W.~de Boer, G.~Burkart and R.~Ehret, \NPB{498}{97}{3}.
%
\bibitem{CEQWHHH} M.~Carena, J.~R.~Espinosa, M.~Quir\'os and C.~E.~M.~Wagner,
\PLB{355}{95}{209};
M.~Carena, M.~Quir\'os and C.~E.~M.~Wagner, \NPB{461}{96}{407};
H.~E.~Haber, R.~Hempfling and A.~H.~Hoang, \ZPC{75}{97}{539}.
%
\bibitem{HH} 
R.~Hempfling and A.~H.~Hoang, \PLB{331}{94}{99}.
%
\bibitem{PBMZ} 
D.~M.~Pierce, J.~A.~Bagger, K.~T.~Matchev and R.-J.~Zhang,
\NPB{491}{97}{3}.
%
\bibitem{HHW} 
S.~Heinemeyer, W.~Hollik and G.~Weiglein,
\PRD{58}{98}{091701}; \PLB{440}{98}{296}; 
\EPJ{9}{99}{343}.
%
\bibitem{RenJie} 
R.-J.~Zhang, \PLB{447}{99}{89}.
%
\bibitem{HHWan} S.~Heinemeyer, W.~Hollik and G.~Weiglein,
\PLB{455}{99}{179}.
%
\bibitem{DR} W.~Siegel, \PLB{84}{79}{19};
D.~M.~Capper, D.~R.~T.~Jones and P.~van~Nieuwenhuizen, \NPB{167}{80}{479};
I.~Jack, D.~R.~T.~Jones, S.~P.~Martin, M.~T.~Vaughn and
Y.~Yamada, \PRD{50}{94}{5481}.
%
\bibitem{FJJ} 
C.~Ford, I.~Jack and D.~R.~T.~Jones,
\NPB{387}{92}{373}, Erratum-ibid. {\bf B504} (1997) 551;
M.~Caffo, H.~Czy\.z, S.~Laporta and E.~Remiddi,
{\it Nuovo Cim.}\ {\bf 111A} (1998) 365.
%
\bibitem{DT}
A.~I.~Davydychev and J.~B.~Tausk, \NPB{397}{93}{123};
A.~I.~Davydychev, V.~A.~Smirnov and J.~B.~Tausk,
\NPB{410}{93}{325};
F.~A.~Berends and J.~B.~Tausk, \NPB{421}{94}{456}.
%
\bibitem{Donini}
A.~Donini, \NPB{467}{96}{3}.
%
\bibitem{martin}
K.~Inoue, A.~Kakuto, H.~Komatsu and S.~Takeshita, \PTP{68}{82}{927}; 
\PTP{71}{84}{413};
S.~P.~Martin and M.~T.~Vaughn, \PRD{50}{94}{2282}; 
Y. Yamada, \PRD{50}{94}{3537};
I. Jack and D.~R.~T.~Jones, \PLB{333}{94}{372}.
%
\end{thebibliography}
\end{document}